# Superconducting nanowire diode

Xiaofu Zhang[1,2,4✉], Qingchang Huan[1,2,3,4], Ruoyan Ma[1,2,4], Xingyu Zhang[1,2], Jia Huang[1,2], Xiaoyu Liu[1,2], Wei Peng[1,2], Hao Li[1,2], Zhen Wang[1,2], Xiaoming Xie[1,2], Lixing You[1,2✉]

[1] National Key Laboratory of Materials for Integrated Circuits, Shanghai Institute of Microsystem and Information Technology, Chinese Academy of Sciences，865 Changning Road, Shanghai 200050 China.
[2] Center of Materials Science and Optoelectronics Engineering, University of Chinese Academy of Sciences, Beijing 100049 China.
[3] School of Microelectronics, Shanghai University, 20 Chengzhong Road, Shanghai 201800 China.
[4] These authors contributed equally to this work.
✉ e-mail: zhangxf@mail.sim.ac.cn; lxyou@mail.sim.ac.cn

**Semiconducting diode with nonreciprocal transport effect underlies the cornerstone of contemporary integrated circuits (ICs) technology. Due to isotropic superconducting properties and the lack of breaking of inversion symmetry for conventional *s*-wave superconductors, such a superconducting peer is absent. Recently, a series of superconducting structures, including superconducting superlattice and quantum-material-based superconducting Josephson junction, have exhibited a superconducting diode effect in terms of polarity-dependent critical current. However, due to complex structures, these composite systems are not able to construct large-scale integrated superconducting circuits. Here, we demonstrated the minimal superconducting electric component-superconducting nanowire-based diode with a nonreciprocal transport effect under a perpendicular magnetic field, in which the superconducting to normal metallic phase transition relies on the polarity and amplitude of the bias current. Our nanowire diodes can be reliably operated nearly at all temperatures below the critical temperature, and the rectification efficiency at 2 K can be more than 24%. Moreover, the superconducting nanowire diode is able to rectify both square wave and sine wave signals without any distortion. Combining the superconducting nanowire-based diodes and transistors, superconducting nanowires hold the possibility to construct novel low-dissipation superconducting ICs.**

The superconducting diode effect[1–9], in terms of Cooper pairs dominated dissipationless supercurrent flowing through the diode in one direction but normal conducting electrons flowing in the opposite direction[9,10], relies on the simultaneous breaking of the inversion and time-reversal symmetry, which is recently realized in a series of composite superconducting systems[1,3-6,11-22]. Therefore, superconducting diodes can be widely applied for rectifiers, *ac–dc* converters, and antennas for detecting electromagnetic signals[12,23,24], especially for extremely low-temperature electronics, where the energy dissipations play dominating roles. However, nearly all the currently realized superconducting diodes consist of rather complex composite structures[11-16,] or quantum materials[1,17-22], and lack the possibility to construct scalable superconducting ICs. To realize the superconducting diode effects in extremely simple architectures with common superconducting materials may remarkably facilitate the application of superconducting diodes and boost the development of superconducting electronics.

Superconducting nanowires, as the minimized functional superconducting electric component, have been successfully demonstrated for fabricating single-photon detectors[25,26] and bipolar transistors[27,28]. In superconducting nanowires, constrictions (either a defect or a sudden expansion) induced current crowding effect and the associated strong phase fluctuations will significantly lower the energy barrier for vortex activation. Subsequently, the activated vortex will cross the whole nanowire driven by the Lorentz force from the bias current[29], underlying the physic origins of dark noise in superconducting nanowire single-photon detectors[30]. Inspired by this mechanism, we designed the asymmetric superconducting nanowire diode (SND) based on constricted nanowires with expanded constrictions. Under a perpendicular magnetic field, the inversion and time-reversal symmetry of the constricted nanowires will be broken simultaneously, realizing the physical foundations for the superconducting nanowire diode.

This article shows that superconducting nanowires with an expanded constriction exhibit a pronounced nonreciprocal transport effect under perpendicular magnetic fields. In particular, the normal to superconducting transition of the constricted nanowires is dependent on the polarity and amplitude of the bias current, presenting distinct absolute $I_c$ values with different bias current polarities. Namely, the superconductivity of the SND relies on the polarities of the bias current and external fields. A superconducting half-wave rectification for both square-wave and sine-wave excitations is demonstrated without distortion. Specifically, the mechanism of SNDs is compatible with nearly all Type-II superconducting nanowires, irrespective of complex architectures, which is highly promising for fabricating universal superconducting diodes. The SNDs, combined with superconducting nanowire transistors (non-sensitive to an external magnetic field), pave the avenue for constructing novel functional superconducting ICs with low dissipations.

The NbTiN nanowire diodes were fabricated on $SiO_2$/Si substrates, as it is shown in Fig. 1a. The SNDs are constituted by a nanowire with a length of 2 μm and a thickness of 8 nm (which is commonly used for fabrications of superconducting nanowire single-photon detectors)[31], and a triangle expansion with a base of 200 nm and a height of 500 nm. Figure 1b shows the zero-field normal to superconducting transition for the 100-nm-wide SND. By fitting the resistance near the transition region with the one-dimensional (1D) superconducting fluctuation theory[32], the superconducting critical temperature $T_c$ is found to be around 9.4 K, and the normal state resistance slightly above $T_c$ exhibits $R_N \sim 10$ kΩ. Moreover, from the field-dependent superconducting transitions, the temperature dependence of the upper critical field can be determined, as shown in detail in Supplementary Fig. S1. The upper critical field in the zero-temperature limit is estimated to be 14.6 T, resulting in a zero-temperature superconducting coherence length of 4.7 nm[33,34].

To investigate the superconducting diode effect, we first measured the *dc* current dependence of the sheet resistance under various magnetic fields and temperatures (Fig. 1c and 1d). Despite the quasi-1D superconducting nature of the constricted nanowires, sharp superconducting to normal transitions are observed at all involved temperatures (Fig. 1c) and magnetic fields (Fig. 1d), therefore allowing for fast switching between the superconducting and normal conducting states. As a typical example of $R(I)$ dependence for $T = 2$ K and $B = 0.26$ T (the solid black line and the short-dotted line in Fig. 1c), the constricted nanowire transitions into the normal state at $I^- = 40.9$ μA and $I^+ = 25.7$ μA, respectively, exhibiting a current polarity dependent superconducting to normal transitions and behaving as an SND[1,2]. Moreover, explicit current polarity-dependent superconducting to normal transitions can also be observed at higher temperatures (Supplementary Fig. S2), indicating the broad operating temperature region of SNDs.

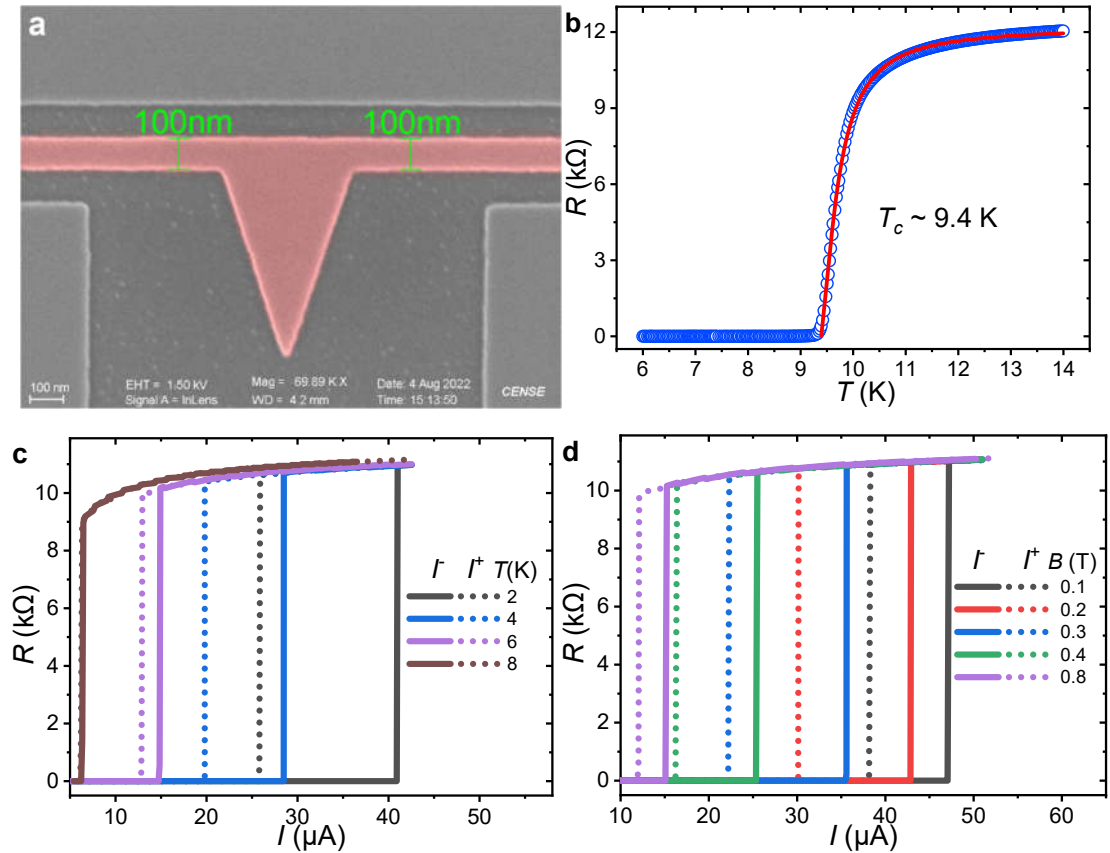

**Fig. 1 | Basic device characterization. a**, False-colour scanning electron micrograph of 100-nm-wide superconducting nanowire diode, with an arrow marking the polarity of bias current. **b**, Resistance as a function of temperature from 14 K to 6 K with an excitation current of 1 μA. The solid red line is a fitting to the one-dimensional superconducting fluctuation theory. **c**, Bias current dependence of the resistance at temperatures from 2 K to 8 K for negative (solid) and positive currents (short dot). The external magnetic field is fixed at 0.26 T. **d**, Field dependent resistance as a function of bias current for negative (solid) and positive currents (short dot). The temperature is set at 2 K throughout the measurement.

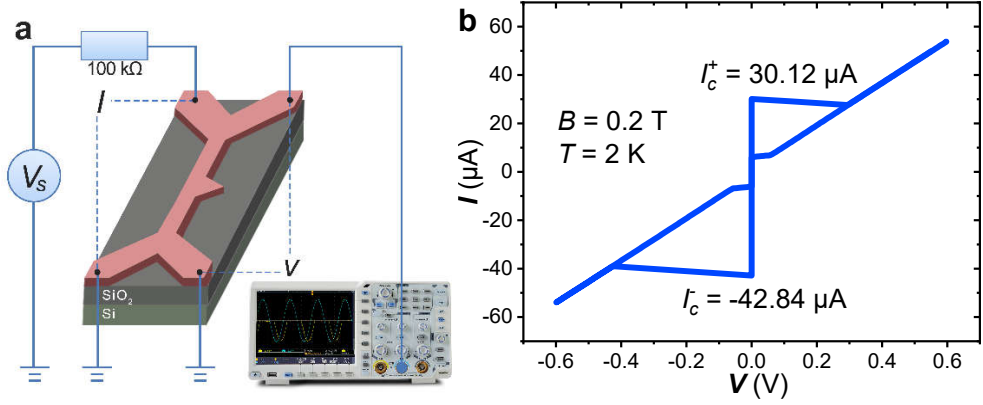

**Fig. 2 | Nonreciprocal transport properties and measurement configuration. a**, Schematic of the measurement configuration and the superconducting nanowire diode. The voltage signal is either collected by a data acquisition card or oscilloscope. **b**, A typical *I-V* characteristic for the SND at $B = 0.2$ T and $T = 2$ K, respectively.

To further reveal the non-reciprocal charge transport in SNDs, we then systematically measured the field-dependent current-voltage (*I-V*) characteristics. Figure 2a schematically shows the measurement setups for the *I-V* characteristics and the rectification effects. The bias current was fed into SNDs from a low-noise voltage source through a 100-kΩ-series resistor. When cooling the SNDs below $T_c$, basically, the Cooper pairs-based dissipationless supercurrent dominates the charge transport in superconducting nanowires. However, under perpendicular magnetic field, due to the breaking of inversion and time-reversal symmetry, the charge transport will be polarity dependent. Figure 2b shows a representative *I-V* characteristic for the 100-nm-SND at a magnetic field and temperature of 0.2 T and 2 K, respectively. Under positive bias current, the SND shows a superconducting to normal transition critical current $I_c^+ = 30.12$ μA. Namely, the charge transport beyond $I_c^+$ will then be dominated by the normal electrons. When bias the SND with negative current, however, the $I_c$ of SND exhibits a significant enhancement to $I_c^- = -42.84$ μA, resulting in a nonreciprocal component $\Delta I_c = -12.72$ μA. As a consequence, when an *ac* current between $I_c^+$ and $I_c^-$ is biased onto the SND, the charge transport in the positive direction is then based on the normal conducting electrons, and dominated by dissipationless Cooper pairs in the opposite direction, underlying the physic foundation for non-reciprocal superconducting diode effect[1,10].

To investigate evolutions of the non-reciprocal superconducting diode effect with the magnetic field, figure 3a presents the *I-V* characteristics for the 100-nm-wide SND under a series of selected external fields at 2 K. All *I-V* curves show evident hysteresis due to the different critical currents $I_c$ for breaking down the superconductivity compared with the returning current $I_r$ for thermalizing the Joule heating. By mapping the superconducting state within the framework of $I_c$ and $B$, an antisymmetric superconducting region emerges in the (*B*, *I*) diagram, which is consistent with previously reported magnetic field involved superconducting diodes[1,15,24]. By extracting the corresponding critical currents from the field-dependent *I-V* characteristics, the polarity dependence of $I_c$ is summarized in Fig. 3b. The maximum point for both $I_c^+(B)$ and $I_c^-(B)$ dependence deviates from the zero-field, indicating the non-homogeneous bias current distribution in the SNDs. Moreover, the nonreciprocal component describing the difference of non-reciprocal charge transport can then be defined as $\Delta I_c = I_c^+ - |I_c^-|$. As it is presented in Fig. 3c, the $\Delta I_c$ is strongly antisymmetric with regard to the magnetic field, which is consistent with the superconductivity mapping shown in Fig. 3a. The nonreciprocal component difference can be further quantified as the rectification efficiency, $\eta = \Delta I_c/(I_c^+ + |I_c^-|)$. As shown in Fig. 3d, the maximum nonreciprocal component difference can be more than 24% at 2 K. To further show the superconducting diode effect at higher temperatures, the magnetic field dependences of nonreciprocal superconducting critical current were also systematically measured at 4.2 K, presenting similar superconducting diode effect but with slightly lower absolute critical current values, as it is shown in detail in Supplementary Fig. S3.

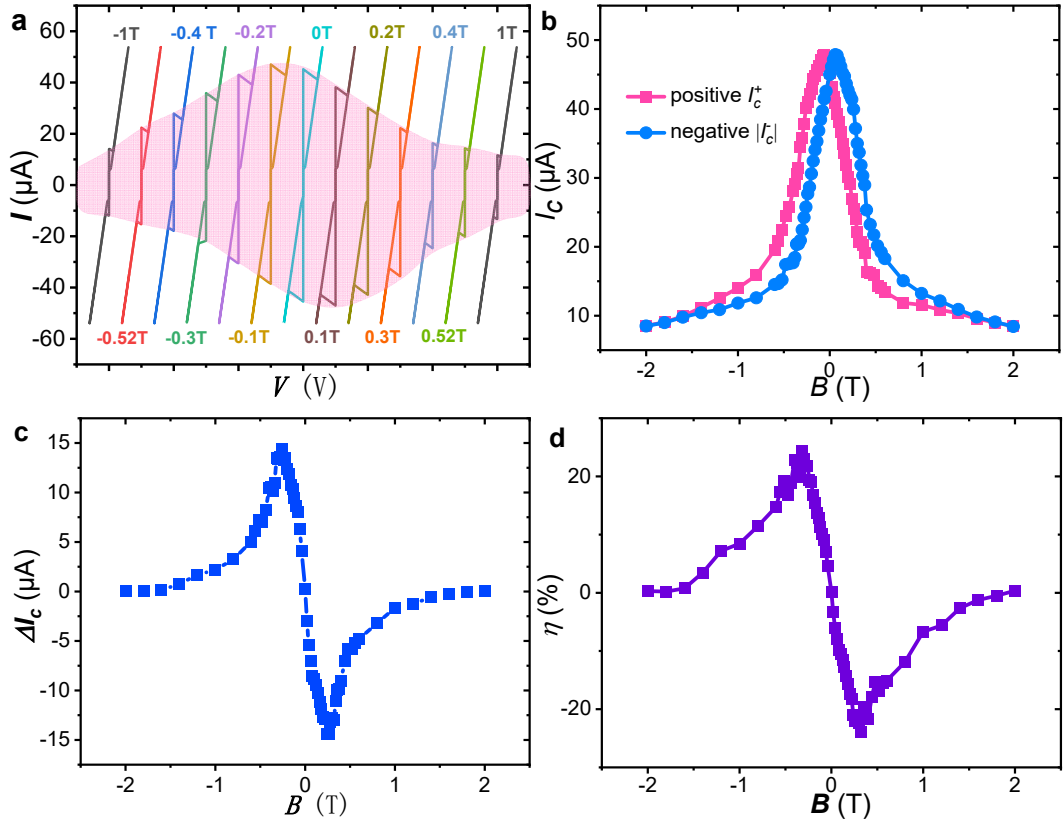

**Fig. 3 | Magnetic field dependences of nonreciprocal superconducting critical current. a**, Magnetic field dependence of current-voltage characteristics for the 100-nm-wide nanowire diode at 2 K. The curves are horizontally offset to show the comparison and evolution of superconducting critical current as a function of the field. The superconducting region is highlighted in the pink region to visualize the asymmetric superconductivity in a $(B, I)$ diagram. **b**, Nonreciprocal critical current $I_c^+$ and $|I_c^-|$ as a function of the magnetic field at 2 K. **c**, Nonreciprocal component of the critical current, $\Delta I_c$, as a function of the magnetic field at 2 K. The maximum $\Delta I_c$ locates at the magnetic field of 0.26 T and -0.26 T, respectively. **d**, The magnetic field dependence of the rectification efficiency, $\eta = \Delta I_c/(I_c^+ + |I_c^-|)$.

Generally, the physical origins of the diode effect in constricted superconducting nanowires can be ascribed to the vortex-related current crowding effect. At zero-field, although the constriction will lower the vortex activation barrier, a similar effect on the anti-vortex from the opposite bias current will counteract the significance of the suppression effect on the critical current from the constriction. This is why the polarity of critical current is absent at zero-field. Under perpendicular fields, for instance, a positive magnetic field and a positive bias current, the Lorentz force on quantized vortices will drive the vortices dome moving towards the constriction-free side, leaving a vortex-free avenue on the constriction edge[34,35]. As a result, the positive bias current mainly flows through the constriction edge, further aggravating the current crowding effect, leading to a strong suppression on $I_c^+$, as shown in Figs. 2b and 3a. On the contrary, when the SND is biased with a negative current, the vortex-free avenue will locate on the constriction-free edge, regardless of the current crowding effect, which subsequently results in larger $I_c^-$ as compared with $I_c^+$ under positive magnetic fields. Based on this mechanism, introductions of asymmetric constrictions with spatial size larger than the phase coherence length to a superconducting nanowire can then realize an SND with a remarkable non-reciprocal superconducting diode effect in a perpendicular external field.

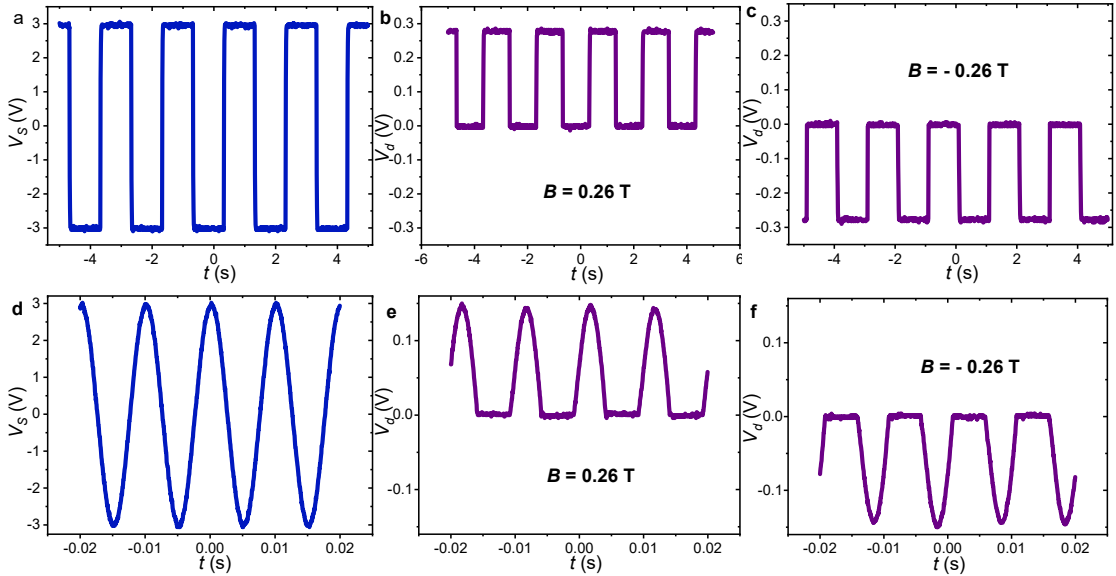

**Fig. 4 | Half-wave rectification for square wave and sine wave signals. a**, The applied square-wave excitation with an amplitude of 3 V, resulting in a 30 μA square-wave bias current on the SND. The length scale of the square-wave is 1 s, and the corresponding frequency is 0.5 Hz. **b,c**, The voltage drop on the SND to the square-wave excitation for $B = 0.26$ T and $-0.26$ T, respectively. **e**, The applied sine-wave excitation of $V = \sqrt{2}V_0 \sin\omega t$, with $V_0 = 2.122$ V and $\omega = 100$ Hz, generating a 30 μA sine-wave bias current on the SND. **e,f**, The dynamic response of the SND to the sine-wave excitation for $B = 0.26$ T and $-0.26$ T, respectively.

Based on the current polarity-dependent superconducting to normal transitions and non-reciprocal charge transport in SNDs, we here further demonstrated the half-wave rectification of SNDs under external fields of 0.26 T and $-0.26$ T, where a maximum nonreciprocal component up to 15 μA is realized, as it is shown in Fig. 3c. According to the nonreciprocal critical current $I_c^+$ and $I_c^-$ shown in Supplementary Fig. S4, a square-wave excitation with an amplitude of 3 V and a frequency of 0.5 Hz (Fig. 4a) was fed onto the SND through the 100-kΩ-series resistor, resulting in a 30 μA square-wave bias current on the SND. For $B = 0.26$ T, since the current amplitude of the square-wave has exceeded the corresponding $I_c^+ = 25.6$ μA, the positive branch of the square-wave will drive the SND into normal state, and voltage drop across the SND is then collected, as it is shown in Fig. 4b. Similarly, for $B = -0.26$ T, the negative part of the square-wave will be blocked by the SNDs, and the collected voltage signal on SNDs is shown in Fig. 4c. Furthermore, we also investigated the half-wave rectification effect of SNDs to sinusoidal current. Figure 4d depicts the applied sine-wave voltage excitation, and the voltage is set to generate a sinusoidal current with a peak value of 30 μA. Notably, the SNDs are able to rectify the sinusoidal current without any distortion (Fig. 4e and 4f). To more explicitly show the rectification effect, the screenshots of the collected signals and the input voltage excitations are presented in Supplementary Fig. S5.

In summary, we have reported the design and fabrication of novel superconducting diodes based on the minimum applicable superconducting electric component, superconducting nanowires. The SND composites consist of short and narrow superconducting nanowire and expanded constrictions with characteristic sizes larger than the corresponding coherence length (a few nanometers) to induce a spatial fluctuation of order parameters. Therefore, the fabrication process of SNDs is rather simple compared to other types of superconducting diodes, fully compatible with the current large-scale IC techniques. Moreover, the SND reported in this work is highly flexible, in which the superconducting properties and the rectification efficiency can be readily adjusted by modifying the characteristic size of nanowire and constriction. By combining the SNDs and superconducting nanowire transistors, it is promising to fabricate novel superconducting logic devices and circuits, paving the way for realizing novel functional superconducting ICs with low dissipations.

Finally, due to the simple operation mechanism, all type-II superconductors can be expected to apply for the fabrications of SNDs. Especially for high $T_c$ superconductors, it is promising to fabricate high $T_c$ SNDs and operate at liquid nitrogen temperature. Moreover, due to the asymmetric vortex energy barriers in SNDs, the SNDs are promising systems for the control of vortex motion and can be further applied for antenna devices to detect environmental fluctuations and broadband electromagnetic radiations[23,24].

**Methods**

**SNDs fabrication.** An 8-nm-thick NbTiN film was first deposited on a 2-inch SiO$_2$/Si substrate by means of magnetron sputtering. The substrate was cleaned with Ar plasma prior to the deposition. Then the designed constricted nanowires were defined by a 100 keV electron beam lithography system using the positive ZEP520 e-beam resist, and the pattern was then transferred by dry etching using reactive ions of CF$_4$ plasma. Afterwards, the contacts for the SNDs are defined by optical lithography, followed by reactive ions etching.

**Transport measurements.** The transport properties of SNDs were measured in a physical property measurement system (PPMS) from Quantum Design. The fabricated SNDs were four-wire connected. The magneto-transport measurements were performed from 2 K to 15 K under various magnetic fields up to 9 T. For measurement in magnetic fields, the field was directed perpendicular to the SND surface.

**Magnetic-field-induced superconducting diode effects.** The non-reciprocal charge transport measurements were performed with a homemade multi-function probe in the PPMS, equipped with a fiber port and four high-frequency electric connections based on coaxial cables. The *dc* voltage excitations are generated by an isolated voltage source (SRS 928, installed with a SIM900 mainframe). The voltage signal on the SNDs is collected with a data acquisition card (NI-9217). The sine-wave voltage excitations are generated by a lock-in amplifier (SR830). The rectified signal and the input waveforms are simultaneously collected with a Tektronix oscilloscope.

**Data availability**

All data supporting the findings of this study are available from the corresponding authors upon reasonable request.

superconducting vortices. Nat. Commun. **8**, 85 (2017).

**Acknowledgements**

We are grateful to X. Liu for the e-beam lithography, L. Zhang and Y. Wu for the deposition of NbTiN films, and W. Zhang for fruitful discussions. This work is sponsored by Shanghai Sailing Program (21YF1455700). The fabrication was performed in the Superconducting Electronics Facility (SELF) of SIMIT, CAS

**Author contributions**

X.Z. and L.Y. supervised this research. X.Z. proposed the idea and conceived the experiments. X.Z., R.M. and Q.H. designed and fabricated the samples. R.M. and Q.H. performed the measurements. All the authors analyzed and interpreted the data and contributed to the writing of the manuscript.

**Competing interests**

The authors declare no competing financial interests.

**Additional information**

**Supplementary information** The online version contains supplementary material available at XXX.

Correspondence and requests for materials should be addressed to X. Zhang or L. You.